\documentclass[conference,10 pt,onecolumn]{IEEEtran}

\IEEEoverridecommandlockouts

\usepackage{graphicx}      
\usepackage[numbers]{natbib}        
\usepackage{graphics} 
\usepackage{epsfig} 
\usepackage{amsmath} 
\usepackage{amssymb}  
\usepackage{txfonts} 
\usepackage{dsfont}
\usepackage{mathrsfs}
\usepackage{graphics}
\usepackage{graphicx}      
\usepackage{caption}
\usepackage{subcaption}
\usepackage{hyperref}
\usepackage{verbatim}
\usepackage{epstopdf}
\usepackage{hyperref}
\usepackage{mathtools}
\usepackage{url}
\usepackage{todonotes}
\usepackage[margin=1.0 in]{geometry}
\usepackage{array}
\usepackage{mathtools}

\usepackage{algpseudocode,algorithm}

\DeclarePairedDelimiter\ceil{\lceil}{\rceil}

\ifCLASSINFOpdf
\else
\fi

\newtheorem{assump}{Assumption}[section]

\newtheorem{prob}{Problem}[section]
\newtheorem{rem}{Remark}[section]

\allowdisplaybreaks[3]

\begin{document}
\title{\Large{\textbf{A New Cyber-Secure Countermeasure for LTI systems under DoS attacks}}}



\author{\IEEEauthorblockN{Nilanjan Roy Chowdhury, Nandini Negi, Aranya Chakrabortty}
\IEEEauthorblockA{Department of Electrical and Computer Engineering, North Carolina State University, USA\\
Email:~nilanjan2008@gmail.com,~nnegi@ncsu.edu,~achakra2@ncsu.edu}}

\maketitle
\begin{abstract}
This paper presents a new counter-measure to mitigate denial-of-service cyber-attacks in linear time-invariant (LTI) systems. We first design a sparse linear quadratic regulator (LQR) optimal controller for a given LTI plant and evaluate the priority of the feedback communication links in terms of the loss of closed-loop performance when the corresponding block of the feedback gain matrix is removed. An attacker may know about this priority ordering, and thereby attack the links with the highest priority. To prevent this, we present a message rerouting strategy by which the states that are scheduled to be transmitted through the high priority links can be rerouted through lower priority ones in case the former get attacked. Since the attacked link is not available for service, and the states of the low priority links can no longer be accommodated either, we run a structured $\mathcal H_2$ control algorithm to determine the post-attack optimal feedback gains. We illustrate various aspects of the proposed algorithms by simulations. 
\end{abstract}

 \ifCLASSOPTIONpeerreview
 \begin{center} \bfseries EDICS Category: 3-BBND \end{center}
 \fi
\IEEEpeerreviewmaketitle

\section{\large{Introduction}}
Security of cyber-physical systems has drawn a significant research attention in recent times. Due to notable instances of cyber-attacks such as WannaCry~\cite{ehrenfeld2017wannacry}, NotPetya~\cite{fayi2018petya}, and Ukranian blackout~\cite{liang20172015}, there have been increased interests to design countermeasures to mitigate attacks for cyber-physical systems. A significant part of the existing work focuses on centralized systems (see e.g.,~\cite{shoukry2016event},~\cite{pasqualetti2013attack}) while the recent results in~\cite{chen2018resilient},~\cite{chen2017distributed} rely on distributed algorithms. Prompted by these considerations, this paper presents a novel sparse optimization based countermeasure to alleviate cyber-attacks for a general class of LTI systems.

System theoretic approaches to tackle a class of cyber-attacks namely Denial of Service (DoS in short) have recently been investigated in~\cite{de2014resilient, de2014non, senejohnny2015self}. Given a LTI system,~\cite{de2014resilient} presents a novel analysis methodology to maintain the closed loop stability under DoS attacks, while~\cite{de2014non} unveils a similar analysis for its nonlinear counterpart. Given a class of complex networks,~\cite{senejohnny2015self} analyzes the consensus property of self-triggered agents in the presence of DoS attacks. The analysis documented in~\cite{senejohnny2015self} introduces the notion of persistence-of-communication and characterizes DoS frequency and duration to attain consensus under DoS attacks.

Substantial studies have been undertaken on the analysis of consensus/ synchronization behavior under DoS attacks, see, for example~\cite{pasqualetti2012consensus, feng2017, feng2017distributed} and the references therein. In the context of multi-agent systems, a group of agents is said to reach consensus/ synchronization when all individuals converge towards a common value (consensus)~/ state (synchronization). Consensus analysis over unreliable networks has been motivated by the seminal contribution from~\cite{pasqualetti2012consensus}, in which, the authors consider a linear consensus model in the presence of misbehaving agents whose behavior diverge from the nominal consensus evolution. Given a complex network having misbehaving agents,~\cite{pasqualetti2012consensus} illustrates the problem of ensuring consensus under non-colluding and Byzantine attacks. Given a class of general LTI systems, consensus under DoS attacks is analyzed for undirected~\cite{feng2017} and directed~\cite{feng2017distributed} topology. The results in~\cite{feng2017, feng2017distributed} hypothesize sufficient conditions to ensure asymptotic consensus and also characterize the frequency and the duration of the DoS interval. However, the above analysis primarily relies on static graphs illustrating an idealistic setup.

Recently game-theoretic results are employed in conjunction with distributed optimization to tackle the security problems for large-scale cyber-physical networks. In the game-theoretic setup, the notion of interdependent security games has recently been explored to compute optimal and strategic security investments by multiple defenders, for example see~\cite{hota2016optimal, hota2018}. In~\cite{hota2016optimal, hota2018}, the authors consider each defender is responsible for the security of multiple assets, in which the inter-dependencies among the assets captured by an interdependency graph. The authors redesign the problem of computing the optimal defense allocation by a single defender as a convex optimization problem and establish the existence of a pure Nash equilibrium of the game between multiple defenders. Given a networked control system,~\cite{shukla2018cyber} investigates a slightly different problem, where the authors reformulate a general-sum, two-player, mixed strategy game between an attacker and a defender. The authors of~\cite{shukla2018cyber} exploit the nonlinear programming paradigm to analyze the dependence of a Mixed Strategy Nash Equilibrium on the relative budgets of the players and preserve important network nodes to attain a desirable LQR performance.\\

\textbf{Summary of Contributions:}~In light of the aforementioned works, in this paper we present a new appoach for mitigating DoS attacks by using ideas from sparse optimal control. Given a LTI system defined over a network of $N\geq2$ nodes, we first design a sparse linear quadratic regulator (LQR) optimal controller using $l_1$-sparsity promotion techniques, proposed in \cite{lin2013design}. The LQR control law is given as $u(t)\,=\,K\,x(t)$. The non-zero blocks $K_{ij}$ of the sparse matrix $K$ indicate the existence of communication links between nodes $i$ and $j$, carrying state $x_i(t)$ to controller $u_j(t)$. We carry out an offline analysis to evaluate the priority of these feedback communication links in terms of how much loss is incurred in the closed-loop LQR objective function when any block $K_{ij}$ is removed. We assume that an attacker may also know about this priority ordering, and thereby is most likely to attack the links with highest priority so that he/she can cause maximum damage to the closed-loop response. To prevent this, we present a message rerouting strategy by which the states that are scheduled to be transmitted through the high priority links can now be quickly rerouted through lower priority ones in case the former becomes dysfunctional from a DoS attack. We present algorithms that capture various practical issues related to the size of the rerouted state vector versus the volume of the low-priority link. One must note that following the re-routing, the attacked link is not available for service, and the states of the low priority links can no longer be accommodated for communication either. Thus, retaining the same control gains for the rest of the states may result in a severely sub-optimal closed-loop performance. We, therefore, finally run a structured $\mathcal H_2$ control algorithm, proposed in \cite{lin2011augmented} to determine the post-attack optimal feedback gains. We illustrate various aspects of the proposed algorithms by simulations. \\

The rest of the paper is organized as follows: In Section~\ref{set} we formulate the problem, while in Section~\ref{pre} we document a preparatory note on the sparsity promoting optimal control problem. We present our proposed rerouting algorithm in Section~\ref{mars}. Finally in Section~\ref{num} we provide an academic example to verify our contribution.\\

\textbf{Notation:}~We denote the set of real and natural numbers by $\mathbb{R}$ and $\mathbb{N}$ respectively. $\mathbf{1}$ is a matrix with all its entries equal to one, while $I_k$ symbolizes the identity matrix of dimension $k$. $\mathbf{0}_{pq}$ denotes a zero matrix of dimension $p\times q$. Given a matrix $M\in\mathbb{R}^{p\times q}$, $||M||_F$ defines the `Frobenius norm' of $M$, and $M(j,:)\in\mathbb{R}^{1\times q}$ presents the $j^{th}$ row of $M$, with $j\in\left\lbrace 1,2,\cdots,p\right\rbrace$. For a square matrix $N\in\mathbb{R}^{p\times p}$, $\text{trace}(N)$ is calculated as $\text{trace}(N)\,:=\,\sum_{i=1}^p n_{ii}$. Given two matrices $A\in\mathbb{R}^{p\times q}$ and $B\in\mathbb{R}^{r\times s}$, $A\otimes B\in\mathbb{R}^{pr\times qs}$ defines their kronecker product. Similarly, for two matrices $C, D\in\mathbb{R}^{m\times n}$, the standard hadamard product is calculated as $C\circ D\in\mathbb{R}^{m\times n}$.

\section{\large{Problem Setup}} \label{set}
Let us consider the continuous-time LTI system as
\begin{align}
\dot{x}(t)\,&=\,Ax(t)\,+\,Bu(t)\,+\,Wd(t), \qquad x(0)\,=\,x_0, \nonumber \\
y(t)\,&=\,Cx(t)\,+\,Du(t), \label{eq1}
\end{align}
where $x(t)\,=\,\begin{bmatrix}
x_1^{\top}(t),\cdots,x_N^{\top}(t)
\end{bmatrix}^\top\in\mathbb{R}^n$ is the overall state vector, with $x_j(t)\in\mathbb{R}^{n_j\times 1}$ is the state corresponding to node $j\in\left\lbrace 1,2,\cdots,N\right\rbrace$, and $n\,=\,\sum_{j=1}^N n_j$, $u(t)\,=\,\begin{bmatrix}
u_1^\top(t),\cdots,u_N^\top(t)
\end{bmatrix}^\top\in\mathbb{R}^{m}$ is the overall control input with $u_j(t)\in\mathbb{R}^{m_j\times 1}$ is the control input of node $j$, and $m\,=\,\sum_{j=1}^N m_j$, $d(t)\in\mathbb{R}^{q}$ is the disturbance, and $y(t)\in\mathbb{R}^p$ is the system output. All the matrices documented above i.e., $A\in\mathbb{R}^{n\times n}, B\in\mathbb{R}^{n\times m}$ and $W\in\mathbb{R}^{n\times q}$ are with appropriate dimensions. In addition, the matrices $C\in\mathbb{R}^{p\times n}$ and $D\in\mathbb{R}^{p\times m}$ are defined as $C:=\begin{bmatrix}
Q^{1/2} & 0\end{bmatrix}^{\top}$ and $D:=\begin{bmatrix}0 & R^{1/2}\end{bmatrix}^{\top}$, where $Q=Q^{\top}\geq 0$ and $R=R^{\top}> 0$ are the state and the control performance weights. We assume, the matrix pair $(A,\,B)$ is stabilizable and $(A,\,Q^{1/2})$ is detectable.

We consider a state feedback control input
\begin{align}
u(t)\,=\,K\,x(t), \label{eq1000}
\end{align}
where $K\in\mathbb{R}^{m\times n}$ is the feedback gain matrix. In the vector form,~\eqref{eq1000} can be expressed as
\begin{align}
\begin{bmatrix}
u_1(t) \\ u_2(t)\\ \vdots \\ u_{N}(t)
\end{bmatrix}\,=\,\begin{bmatrix}
K_{11} & K_{12} & \cdots & K_{1N}\\
K_{21} & K_{22} & \cdots & K_{2N}\\
\vdots & \vdots & \ddots & \vdots \\
K_{N1} & K_{N2} & \cdots & K_{NN}
\end{bmatrix}\,\begin{bmatrix}
x_1(t) \\ x_2(t)\\ \vdots \\ x_{N}(t)
\end{bmatrix},
\end{align}
where the sub-matrix $K_{ij}\in\mathbb{R}^{m_i\times n_j}$ represents a communication link that delivers a block of information of state $j\in\left\lbrace 1,\cdots, N\right\rbrace $ to control $i\in\left\lbrace 1,\cdots,N\right\rbrace$. For simplicity, for $i\,=\,j$, the feedback links are referred to as \textit{local links}, while for $i\,\neq\,j$ as \textit{communication links}.

Notice that, given a LTI system~\eqref{eq1}, an optimal gain $K\in\mathbb{R}^{m\times n}$ can be designed employing the LQR strategy~\cite{zhou1996robust}.\\ 
When $K$ is optimal, every $K_{ij}$ and $K_{ii}$, in general, are non-zero sub-matrices, implying that the communication network required for exchanging the states is a dense graph i.e., communication links from every state to every control input. Such dense graphs can result in high communication costs. Therefore, to reduce the cost, we impose an additional structural constraint $\Omega$ on the structure of $K$ as follows~\cite{lin2011augmented}
\begin{equation}
\begin{aligned}
& \underset{K}{\text{minimize}}
& & J(K) \\
& \text{subject to}
& & K\,\in\,\Omega,
\end{aligned}      \label{eq3}
\end{equation}
where the cost function $J(K)$ is defined as
\[ J(K)\,:=\,\begin{cases}
      \text{trace}\,(W^{\top}P(K)W) & \text{if $K$ is stabilizing} \\
       +\infty & \text{otherwise}.
      \end{cases}
\]
The $P(K)\in\mathbb{R}^{n\times n}$ matrix stated above, denotes the closed-loop observability Gramian and presented as
\begin{align}
P(K)\,:=\,\int_0^\infty e^{(A-BK)^\top\sigma} (Q\,+\,K^{\top}RK)e^{(A-BK)\,\sigma}\, d\sigma,
\end{align}
which is obtained by solving the Lyapunov equation 
\begin{align}
(A-BK)^{\top}P\,+\,P\,(A-BK)\,=\,- (Q\,+\,K^{\top}RK).
\end{align}

Given a system~\eqref{eq1} and~\eqref{eq1000}, let $K_1\in\Omega$ be the solution of~\eqref{eq3}, in which, a communication link $K_{1_{ij}}\in\mathbb{R}^{m_i\times n_j}$ delivers a block of information of state $j$ to control $i$. We assume, during the closed loop operation, an attacker attacks either a communication link or a local link of $K_1$ to destabilize the system. For instance, killing a communication/ local link say $K^*_{1_{ij}}$, containing $r\,>\,0$, number of messages equivalently signifies $K^*_{1_{ij}}$ is zeroed out. Let $K_2\in\mathbb{R}^{m\times n}$ be the post-attack feedback gain in which $K^*_{1_{ij}}\,=\,\mathbf{0}_{ij}$. Therefore, after the attack, we focus on addressing the objective of:\\
\begin{prob} \label{pro1}
\textcolor{blue}{finding some communication space in $K_2$ to reroute the attacked $r$ messages. In particular, we need to determine `$r$ spots' in the off-diagonal and the diagonal blocks of $K_2$ such that the resultant closed loop system is stable, and it minimizes the cost function $J(K_2)$.}
\end{prob} 
\section{\large{Technical preliminaries}} \label{pre}
In the following, we briefly review the sparsity promoting optimal control problem. The readers are encouraged to see~\cite{lin2013design} for details and further references.

\subsection{Sparsity-Promoting Optimal Control Problems} \label{adm}
The optimization problem~\eqref{eq3} solely relies on the structure of the communication graph. Hence, in the following, we characterize the optimization setup, in which the sparsity of the feedback gain directly subsumed into the objective function as
\begin{equation}
\begin{aligned}
& {\text{minimize}}
& & J(K)\,+\,\beta\,card\,(K), \\
\end{aligned}      \label{eq4}
\end{equation}  
where, $card: \mathbb{R}^{m\times n}\rightarrow \mathbb{N}$ is the cardinality function i.e., the number of nonzero elements of a matrix (denoted as \,$nnz\,(\cdot)$), and is defined as:
\begin{align}
card\,(K)\,:=\,\sum_{i,j=1}^N nnz\,(K_{ij}).
\end{align}
In addition,~$\beta\,\in\,\left[0,\infty\right)$ is a scalar gain, and a large value of $\beta$ leads to a sparser $K$.\\

Since the objective function~\eqref{eq4} is non-convex due to the cardinality function, therefore it is typically replaced by the weighted $l_1$ norm of the optimization variables~\cite{lin2013design}. Given a feedback gain $K\in\mathbb{R}^{m\times n}$, the weighted $l_1$ norm is represented as
\begin{align}
card\,(K)\,=\,\sum_{i,j}\,G_{ij}\,||K_{ij}||_F, \label{eq6}
\end{align}
where  $G_{ij}$ are the non-negative weights, defined as $G_{ij}\,:=\,1/\left(||K_{ij}||_F+\varepsilon\right)$ with $0\,<\,\varepsilon\,\ll\,1$. Therefore, the minimization problem~\eqref{eq4} can further be approximated as
\begin{equation}
\begin{aligned}
& {\text{minimize}}
& & J(K)\,+\,\beta\,\sum_{i,j}\,G_{ij}\,||K_{ij}||_F, \\
\end{aligned}      \label{eq5}
\end{equation}  
where, $J(K)$ is the square of the closed loop $\mathcal{H}_2$ norm and $\sum\limits_{i,j}\,G_{ij}\,||K_{ij}||_F$ is the sparsity promoting penalty function.

\section{\large{Proposed Rerouting Algorithm}}\label{mars}
Given a continuous-time system~\eqref{eq1}, this section presents a rerouting algorithm to mitigate the cyber-attacks for LTI systems. Our strategy can typically be categorized in three steps : i)~\textbf{link prioritization ranking algorithm}, ii)~\textbf{rerouting algorithm}, and iii)~\textbf{structured $\mathcal{H}_2$ algorithm}. \\

\begin{rem} \label{rema123}
{\rm As indicated earlier, we consider $K$ in~\eqref{eq1000} to be partitioned into $r_1$ non-zero block matrices. Each block $K_{ij}$ is associated with a communication link carrying state $x_i(t)$ from the $i^{th}$ node to controller $u_j(t)$ at the $j^{th}$ node. When an attacker deactivates a link, the entire block $K_{ij}$ is zeroed out. Let us assume the block $K_{ij}$ contains $r$ number of messages. Hence, as defined in Problem~\ref{pro1}, we need ``$r$ spots" in the lower priority control channels to reroute the attacked messages. Now, let us consider a scenario, in which, we do not have a lower priority control channel to reroute the entire $r$ messages. For instance, we assume the lower priority control channels can reroute only $\hat{r}\,(<\,r)$ unit of information. Under this circumstance, we further partition the attacked $r$ messages and use $\ceil*{\frac{r}{\hat{r}}}$ communication channels to reroute the entire message.}
\end{rem}
\vspace{0.2 cm}
\textbf{Step~$1$: Link prioritization ranking algorithm:}~Given a LTI system~\eqref{eq1} with the feedback gain $K\in\mathbb{R}^{m\times n}$, this step evaluates the priority of the feedback communication links $K_{ij}$. It typically carries out an offline analysis technique to evaluate the priority of the feedback communication links in terms of how much loss is incurred in the closed-loop LQR objective function when any block $K_{ij}$ is removed. We initialize a small value of $\beta\in\mathbb{R}$ denoted as $\beta_{\text{initial}}$. It results a centralized LQR gain $K\,=\, K_c\in\mathbb{R}^{m\times n}$, and starts minimizing~\eqref{eq5}. As stated earlier, the solution of~\eqref{eq5} becomes sparser as $\beta$ increases; hence we obtain different footprints of the sparsity pattern of $K$ by varying $\beta$. Therefore, from these different sparsity patterns of $K$, we accumulate the knowledge of the priority of each control blocks. Following this, we construct a matrix $N_{\text{initial}}:\mathbb{R}^{m\times n}\rightarrow \mathbb{R}^{r_1\times r_2}$ in which the control blocks are placed based on their priority. In the sequel, we denote $r_1\in\mathbb{N}$ as the total number of non-zero control blocks present in $K$, while $r_2\in\mathbb{N}$ defines the `size' of each block i.e. the amount of information stored in each block.
\begin{algorithm}[H] 
\caption{(Offline) link prioritization ranking algorithm}
\begin{algorithmic}[1]
\State \textbf{Input data:} $A, B, W, Q, R$ from~\eqref{eq1}, and $\beta_{\text{initial}}$,
\State \textbf{Output data:} $K$ and $N_{\text{initial}}(K)$,
\State\vspace{0.1 cm} \textit{Initialize} $ \beta_1\,=\,\beta_{\text{initial}}$, $N_o(\cdot)\,=\,\mathbf{0}_{r_1\,r_2}$,
\For {$j\,=\,\beta_1\,,\,\beta_2\,,\,\beta_3\,,\,\cdots$~with,~$\beta_1\,<\,\beta_2\,<\,\beta_3\,<\,\cdots$}
\State Minimize~\eqref{eq5} and obtain the total number of nonzero control blocks in $K$ at $\beta_j$,
\State Determine the control blocks which are vanished to sparsify $K$ at $\beta_j$ w.r.t $\beta_{j-1}$,
\State Assign the vanished control blocks in $N_o(\cdot)$, with the priority pertaining to $\beta_j$,
\EndFor
\State \textit{Return} $N_{\text{initial}}(K)\,=\,N_o(\cdot)$.
\end{algorithmic}
\end{algorithm}

\begin{figure}[!htb]
	\centering
		\includegraphics[width = 3.0 in, keepaspectratio]{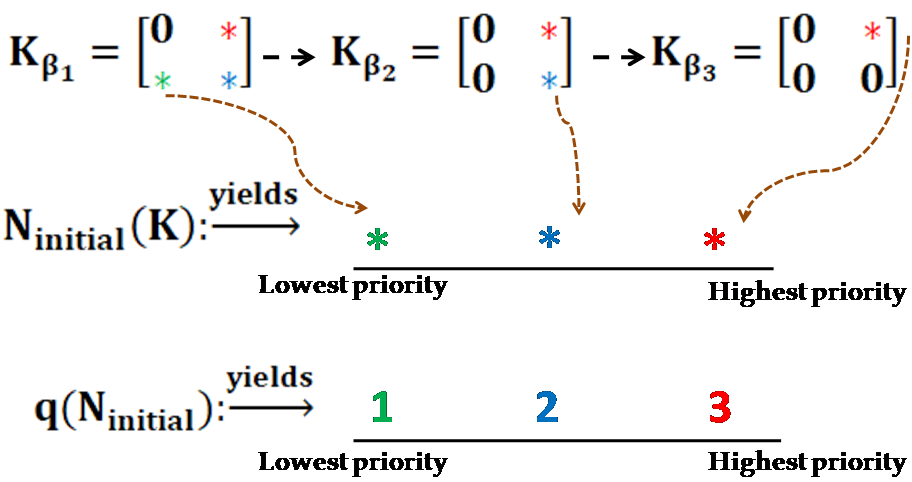}
	\caption{\small{A pictorial representaion of Algorithm~$1$ with $\beta_1<\beta_2<\beta_3$.}} \label{1211}
\end{figure}

The functionality of Algorithm~$1$ is illustrated in Fig.~\ref{1211}.

\begin{rem}
The \textit{link prioritization ranking} algorithm stated above, is performed offline, and it typically considers the class of problems in which the attacked communication links do not vary with time. For instance, let us assume $\mathcal{S}(t^*)$ is the set of communication channels which are attacked by the attacker at time $t^*\,\geq\,0$, then the \textit{link prioritization ranking} algorithm considers $\mathcal{S}(t)\,=\,\mathcal{S}(t^*)$ for all $t\,\geq\,t^*$.   
\end{rem}

\textbf{Step~$2$: Rerouting\footnote{In this work, the notion of rerouting signifies to give up some lower priority communication links to sustain an attacked one.} algorithm:}~Given a continuous-time LTI system~\eqref{eq1} and~\eqref{eq1000}, Step-$1$ generates $N_{\text{initial}}(K)\in\mathbb{R}^{r_1\times r_2}$ accumulating all the non-zero control blocks of $K\in\mathbb{R}^{m\times n}$ and assigns priority to each of them. In addition, let $q: \mathbb{R}^{r_1\times r_2}\rightarrow \mathbb{R}^{r_1}$ be the \textit{priority vector}, which hold the priority information of the control blocks stacked in $N_{\text{initial}}(K)$. We assume the attacker knows $N_{\text{initial}}(K)$ and $q(N_{\text{initial}}(K))\in\mathbb{R}^{r_1}$ and based on this information, it attacks communication links to destabilize the overall closed-loop system. Let $p_{\text{attack}}(\cdot)\in\mathbb{R}^{r_1}$ be the \textit{attack index}, and it contains information about the attacked control blocks/ communication links, in which, the $j^{th}$ row denoted as $p_{\text{attack}}(j)$, defined as:
\[ p_{\text{attack}}(j)\,:=\,\begin{cases}
      1 & \text{if $N_{\text{initial}}(j,:)$ is under attack}, \\
      0 & \text{if $N_{\text{initial}}(j,:)$ is not attacked}.
      \end{cases}
\]
To design a countermeasure, we assume the defender knows $N_{\text{initial}}(K)$, $q(\cdot)$ and $p_{\text{attack}}(\cdot)$. In the following, we present three variants of the rerouting algorithms described in Algorithm~$2- 4$. \\

Let us remark, the first rerouting algorithm given in Algorithm~$2$, based on the following assumption:\\

\begin{assump} \label{assume2}
All the sub-matrices $K_{ij}$ of $K$, are with the same dimension.
\end{assump} 
\vspace{0.2 cm}
Given a feedback gain matrix $K$, the above assumption illustrates that all the block matrices $K_{i\,j}\in\mathbb{R}^{m_{i}\times n_j}$ are required to construct $K$, having same dimension. Additionally, it also signifies that all the state vectors in~\eqref{eq1}, having same dimension, i.e., $x_j(\cdot)\in\mathbb{R}^{n_j}$ where, $n_1=n_2=\cdots=n$ for all, $j\in\left\lbrace 1,2,\cdots,N\right\rbrace$.
 
\begin{algorithm}[H]\label{al}
\caption{Rerouting algorithm: all the sub-matrices have \textbf{same} dimensions}
\begin{algorithmic}[1]
\State \textbf{Input data:} $N_{\text{initial}}(K),~q(\cdot)$, and $p_{\text{attack}}(\cdot)$,
\State \textbf{Output data:} $N_{\text{final}}(K)$,
\State\vspace{0.1 cm} \textit{Initialize} $\mathbb{R}^{r_1\times r_2}\ni N(\cdot)\,:=\,N_{\text{initial}}(K)$,
\State Determine the attacked links from $p_{\text{attack}}(\cdot)\in\mathbb{R}^{r_1}$ and the corresponding priority of the attacked links are stored in $a^*(\cdot)\in\mathbb{R}^{r_3}$, where $r_3\,\leq\,r_1$, 
\State Sort $a^*(\cdot)$ in \textit{descending order} and denote as $a1(\cdot)\in\mathbb{R}^{r_3}$,
\If{($r_3\,=\,0$)}
\State STOP. No attack has occurred.
\ElsIf {($r_3\,>\,r_1/2$)}
\State STOP. Countermeasure can not be implemented.
\Else
\For{($j\,=\,1,\,\,2,\,\cdots,\,r_3$)} 
\If{$(p_{\text{attack}}(j)\,=\,1)~~\&~~(a1(j)\,<\,q(j+1))$}
\State\vspace{0.1 cm} Lower priority communication space is unavailable, discard the attacked control block information by $N(a1(j),:)\,=\,\mathbf{0}_{1\,r_2}$,
\Else
\State\vspace{0.1 cm} Lower priority communication space is available, reroute high priority information by $N(j,:)=\mathbf{0}_{1\,r_2}$,
\EndIf 
\EndFor
\EndIf
\State \textit{Return} $N_{\text{final}}(K)\,=\,N(\cdot)$.
\end{algorithmic}
\end{algorithm}

We present the following example to demonstrate Algorithm~$2$:\\

\textit{Example~$1$:} Consider a continuous-time LTI system~\eqref{eq1} with $(A,B)\in\mathbb{R}^{8\times 8}\times \mathbb{R}^{8\times 4}$. Let $K\in \Omega_1$ be the solution of~\eqref{eq3}. The structural constraint $\Omega_1\in \mathbb{R}^{4\times 8}$ is given as
\begin{align*}
\Omega_1\,=\,\begin{bmatrix}
\textcolor{cyan}{\star} & \mathbf{0} & \textcolor{cyan}{\star} & \textcolor{cyan}{\star}\\
\mathbf{0} & \textcolor{cyan}{\star} & \mathbf{0} & \textcolor{cyan}{\star}\\
\mathbf{0} & \textcolor{cyan}{\star} & \mathbf{0} & \mathbf{0}\\
\textcolor{cyan}{\star} & \textcolor{cyan}{\star} & \mathbf{0} & \textcolor{cyan}{\star}
\end{bmatrix},
\end{align*}
where $\textcolor{cyan}{\star}$ and $\mathbf{0}$ denote a nonzero and a zero matrices of dimension $1\times 2$ respectively. Let $K\in\Omega_1$ be evaluated as
\begin{align}
K\,=\,\begin{bmatrix}
\textcolor{cyan}{3} & \textcolor{cyan}{1} & 0 & 0 & \textcolor{cyan}7 & \textcolor{cyan}9 & \textcolor{cyan}3 & \textcolor{cyan}2\\
0 & 0 & \textcolor{cyan}1 & \textcolor{cyan}5 & 0 & 0 & \textcolor{cyan}1 & \textcolor{cyan}2\\
0 & 0 & \textcolor{cyan}5 & \textcolor{cyan}1 & 0 & 0 & 0 & 0\\
\textcolor{cyan}2 & \textcolor{cyan}4 & \textcolor{cyan}6 & \textcolor{cyan}8 & 0 & 0 & \textcolor{cyan}5 & \textcolor{cyan}3
\end{bmatrix}. \label{eq12345}
\end{align}
Given a feedback gain $K$ stated above, we employ algorithm $1$ to evaluate the priorities of the non-zero blocks $K_{ij}$. The priorities corresponding to the non-zero blocks $K_{ij}$ are presented as $q(\cdot)=1\mapsto \begin{bmatrix} \textcolor{cyan}{3} & \textcolor{cyan}{1} \end{bmatrix}$, $q(\cdot)=2\mapsto \begin{bmatrix} \textcolor{cyan}{2} & \textcolor{cyan}{4} \end{bmatrix}$, $q(\cdot)=3\mapsto \begin{bmatrix} \textcolor{cyan}{1} & \textcolor{cyan}{5} \end{bmatrix}$,
$q(\cdot)=4\mapsto \begin{bmatrix} \textcolor{cyan}{5} & \textcolor{cyan}{1} \end{bmatrix}$,
$q(\cdot)=5\mapsto \begin{bmatrix} \textcolor{cyan}{6} & \textcolor{cyan}{8} \end{bmatrix}$,
$q(\cdot)=6\mapsto \begin{bmatrix} \textcolor{cyan}{7} & \textcolor{cyan}{9} \end{bmatrix}$,
$q(\cdot)=7\mapsto \begin{bmatrix} \textcolor{cyan}{3} & \textcolor{cyan}{2} \end{bmatrix}$,
$q(\cdot)=8\mapsto \begin{bmatrix} \textcolor{cyan}{1} & \textcolor{cyan}{2} \end{bmatrix}$,
$q(\cdot)=9\mapsto \begin{bmatrix} \textcolor{cyan}{5} & \textcolor{cyan}{3} \end{bmatrix}$.

\begin{figure}[!htb]
	\centering
		\includegraphics[width = 3.5 in, keepaspectratio]{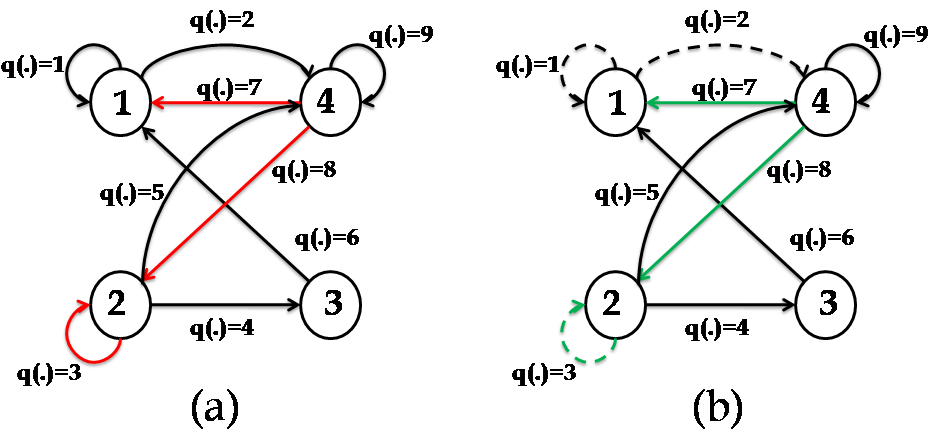}
	\caption{\small{(a) Communication topology of the control gain $K\in\Omega_1$ given in~\eqref{eq12345}, where ($\textcolor{red}{\bullet}$) defines the links which are attacked, (b) post-attack communication graph in which ($\textcolor{green}{\bullet}$) denotes the links whose data safely rerouted through lower priority blocks. The dotted line signifies lower priority communication links which are sacrificed to reroute the attacked information. Here each communication link carries $2$ unit of information, while the priority of each link denoted as $q(\cdot)$, are specified over it.}} \label{121212}
\end{figure}

We assume the attacker knows the priorities of the control blocks $K_{ij}$ stated above, and it attacks the control blocks corresponding to the priority level $q(\cdot)=3, 7, 8$. Then, employing Algorithm~$2$, we obtain the post-attack $N_{\text{final}}(K)$ as
$q(\cdot)=1\mapsto \begin{bmatrix} 0 & 0 \end{bmatrix}$, $q(\cdot)=2\mapsto \begin{bmatrix} 0 & 0 \end{bmatrix}$, $q(\cdot)=3\mapsto \begin{bmatrix} \textcolor{green}{0} & \textcolor{green}{0} \end{bmatrix}$,
$q(\cdot)=4\mapsto \begin{bmatrix} 5 & 1 \end{bmatrix}$,
$q(\cdot)=5\mapsto \begin{bmatrix} 6 & 8 \end{bmatrix}$,
$q(\cdot)=6\mapsto \begin{bmatrix} 7 & 9 \end{bmatrix}$,
$q(\cdot)=7\mapsto \begin{bmatrix} \textcolor{green}{3} & \textcolor{green}{2} \end{bmatrix}$,
$q(\cdot)=8\mapsto \begin{bmatrix} \textcolor{green}{1} & \textcolor{green}{2} \end{bmatrix}$,
$q(\cdot)=9\mapsto \begin{bmatrix} 5 & 3 \end{bmatrix}$.\footnote{The null matrix $\begin{bmatrix}0 & 0 \end{bmatrix}$ represented as the dotted line in Fig.~\ref{121212}, signifies the control blocks which are scarified to reroute the attacked information.}\\

Notice that, the defender reroutes the attacked information with priority level $q(\cdot)\,=\,8$ and $q(\cdot)\,=\,7$ via the lower priority control channels with priority $q(\cdot)\,=\,1$ and $q(\cdot)\,=\,2$ respectively (see Step~$15$ in Algorithm~$2$). Likewise, an alternative communication link to redirect the attacked information of the control block with $q(\cdot)\,=\,3$ will be the control block with $q(\cdot)\,=\,4$. However, the priority of the control block with $q(\cdot)\,=\,4$ is higher than the priority of the control block with $q(\cdot)\,=\,3$. Hence the defender discards the information corresponding to $q(\cdot)\,=\,3$ (see Step~$13$ in Algorithm~$2$); see Fig.~\ref{121212} for an illustration.\\


It is noteworthy that Algorithm~$2$ is designed based on Assumption~\ref{assume2}, which implies each control block contains the same amount of information. In the subsequent algorithms, we relax Assumption~\ref{assume2} and consider a more generalized setup, in which, each control block contains different amount of information. Hence in the sequel, we present two variants of the rerouting algorithms described in Algorithm~$3$ and $4$, in which Algorithm~$3$ presents a rerouting strategy for a single link attack, while Algorithm~$4$ elucidates multiple links attack.\\

\begin{rem}
{\rm As stated in Step~$1$, given an $N_{\text{initial}}(\cdot)\in\mathbb{R}^{r_1\times r_2}$, $r_1$ denotes the total number of the non-zero control blocks present in the gain matrix $K$, while $r_2$ defines the size of each block. However, Algorithm $3$ and $4$ consider a $K$ matrix, in which, each control blocks are with different sizes. Let, $\bar{r}_1, \bar{r}_2,\cdots, \bar{r}_n$ with $n\in\mathbb{N}$, be the sizes of the control blocks present in $K$, then in the sequel, we select $r_2\,:=\,\max\left\lbrace \bar{r}_1, \bar{r}_2,\cdots, \bar{r}_n\right\rbrace$.}
\end{rem}
\vspace{0.2 cm}
Subsequently in the Algorithm~$3$, $r_{\text{attack}}$ denotes the single link which is attacked. 

\begin{algorithm}[H]\label{a2}
\caption{Rerouting algorithm: non-zero control blocks of different sizes with a \textbf{single link} attack}
\begin{algorithmic}[1]
\State \textbf{Input data:} $N_{\text{initial}}(K),~r_{\text{attack}}$,
\State \textbf{Output data:} $N_{\text{final}}(K)$,
\State \vspace{0.1 cm} \textit{Initialization} $\mathbb{R}^{r_1\times r_2}\ni N(\cdot)\,:=\,N_{\text{initial}}(K)$,
\State Define $s(\cdot)\in\mathbb{R}^{r_1}$, contains the amount of information accumulated by each control block,
\If {($r_{\text{attack}}\,=\,1$)}
\State Set $N(r_{\text{attack}},:)\,=\,\mathbf{0}_{1\,r_2}$,
\Else
\State Set $c_2\,=\,s(r_{\text{attack}})$ and $c_1\,=\,0$,
\State Calculate the number of available (not attacked) communication spaces (denoted as $c_1\in\mathbb{R}$) to reroute $c_2\in\mathbb{R}$ attacked messages,
\If {($c_1\,<\,c_2$)} 
\State Insufficient lower priority communication spaces are available; set $N(r_{\text{attack}},:)\,=\,\mathbf{0}_{1\,r_2}$,
\Else
\State Set $k\,=\,0$,
\While {($c_2\,>\,0$)}
\State Update $k\,=\,k\,+\,1$ and reroute $c_2$ information via low priority control channel $N(k,:)\,=\,\mathbf{0}_{1\,r_2}$,
\State Update $c_2\,=\,c_2\,-\,s(k)$,
\EndWhile
\EndIf
\EndIf
\State \textit{Return} $N_{\text{final}}(K)\,=\,N(\cdot)$.
\end{algorithmic}
\end{algorithm}

We present a variant of Algorithm $3$ with multiple links attack, as follows:

\begin{algorithm}[H]\label{a3}
\caption{Rerouting algorithm: non-zero control blocks with different sizes with \textbf{multiple links} attack}
\begin{algorithmic}[1]
\State \textbf{Input data:} $N_{\text{initial}}(K),~p_{\text{attack}}(\cdot)$,
\State \textbf{Output data:} $N_{\text{final}}(K)$,
\State \vspace{0.1 cm} \textit{Initialization} see Step~$3$ of Algorithm $3$.
\State Set $b_1\,=\,b_2\,=\,0$,
\State Calculate $s(\cdot)\in\mathbb{R}^{r_1}$ given in Step $4$ of Algorithm $3$,
\State Calculate $a^*(\cdot)\,,\,a1(\cdot)\in\mathbb{R}^{r_3}$ given in Step $4-5$ of Algorithm $2$,
\State Determine the location of the attacked control block with the highest priority and denote as $a^*(\text{end})\in\mathbb{R}$,
\State\vspace{0.1 cm} Calculate the total number of attacked information (determined as $b_1\in\mathbb{R}$) and the available communication space to reroute the attacked $b_1$ information (determined as $b_2\in\mathbb{R}$) upto $a^*(\text{end})$,
\If {($b_1\,=\,0$)}
\State STOP. No attack has happened.
\ElsIf {$(b_2\,=\,0)~~\&~~(r_3\,<\,r_1/2)$}
\For {($j\,=\,1\,,\cdots,\,r_3$)}
\State Set $N(j,:)\,=\,\mathbf{0}_{1\,r_2}$,
\EndFor
\ElsIf {$(b_1\,>\,b_2)~~\&~~(r_3\,\geq\,r_1/2)$}
\State STOP. Countermeasure can not be designed.
\Else
\For {($j\,=\,1\,\cdots\,r_3$)}
\State Set $c_2\,=\,s(a1(j))$ and $c_1\,=\,0$,
\State Go to Step $8\,-\,19$ of Algorithm $3$,
\EndFor
\EndIf
\State \textit{Return} $N_{\text{final}}(K)\,=\,N(\cdot)$.
\end{algorithmic}
\end{algorithm}

\begin{figure}[!htb]
	\centering
		\includegraphics[width = 3.5 in, keepaspectratio]{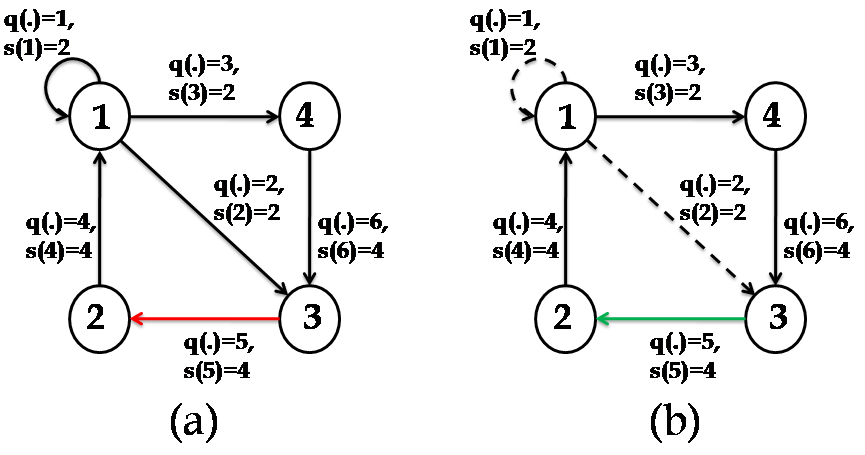}
	\caption{\small{(a) Communication topology of $K\in\Omega_2$ given in~\eqref{eq123456}, where ($\textcolor{red}{\bullet}$) signifies the attacked link, (b) post-attack communication graph where ($\textcolor{green}{\bullet}$) denotes the link whose data are safely rerouted through lower priority blocks. The dotted line denotes lower priority communication links which are sacrificed to reroute the attacked information, while, the nomenclature stated above the communication links e.g., $s(\cdot)$ and $q(\cdot)$, specifies the amount of information communicated over that links and the priority of the link respectively.}}\label{121213} 
\end{figure}

\begin{rem}
{\rm Notice that, in the algorithmic setup given in Algorithm $3$ and $4$, we recall the class of gain matrices $K$ in which, each communication link (say $K_{ij}$) carries different amounts of information corresponding to state $x_i(t)$ from the $i^{th}$ node to controller $u_j(t)$ at the $j^{th}$ node. Under these circumstances, a single link attack describes a scenario, where the attacker attacks only one communication link, while in a multiple links attack, the attacker kills several (more than one) communication channels. However, Algorithm $4$ is capable enough to tackle multiple links attack problem; it can also be implemented for a single link attack case as well. In other words, Algorithm~$3$ is considered a simpler version of Algorithm $4$ and provides a much-needed foundation to design Algorithm~$4$.}
\end{rem}
\vspace{0.2 cm}
\textit{Example~$2$:}~This example illustrates the functionality of Algorithm $4$. We revisit the continuous time LTI system~\eqref{eq1} with $(A, B)\in\mathbb{R}^{14\times 14}\times \mathbb{R}^{14\times 4}$. Let $K\in\Omega_2$ be the solution of~\eqref{eq3}, where $\Omega_2$ is structured as
\begin{align}
\Omega_2\,=\,\begin{bmatrix}
\textcolor{cyan}{\star} & \textcolor{blue}{\star} & \mathbf{0} & \mathbf{0} \\
\mathbf{0} & \mathbf{0} & \textcolor{blue}{\star} & \mathbf{0} \\
\textcolor{cyan}{\star} & \mathbf{0} & \mathbf{0} & \textcolor{blue}{\star} \\
\textcolor{cyan}{\star} & \mathbf{0} & \mathbf{0} & \mathbf{0} 
\end{bmatrix}. \label{eq4001}
\end{align}  
$\textcolor{cyan}{\star}$ and $\textcolor{blue}{\star}$ denote non-zero control blocks having dimensions $1\times 2$ and $1\times 4$ respectively. In other words, it contains $2$ and $4$ unit of information. The feedback gain $K$ is evaluated as
\begin{align}
K\,=\,
\begin{bmatrix*}[r]
\textcolor{cyan}{2} & \textcolor{cyan}1 & \textcolor{blue}3 & \textcolor{blue}7 & \textcolor{blue}5 & \textcolor{blue}8 & 0 & 0 & 0 & 0 & 0 & 0 & 0 & 0\\
0 & 0 & 0 & 0 & 0 & 0 & \textcolor{blue}3 & \textcolor{blue}1 & \textcolor{blue}3 & \textcolor{blue}6 & 0 & 0 & 0 & 0\\
\textcolor{cyan}1 & \textcolor{cyan}5 & 0 & 0 & 0 & 0 & 0 & 0 & 0 & 0 & \textcolor{blue}7 & \textcolor{blue}2 & \textcolor{blue}6 & \textcolor{blue}4\\
\textcolor{cyan}3 & \textcolor{cyan}5 & 0 & 0 & 0 & 0 & 0 & 0 & 0 & 0 & 0 & 0 & 0 & 0
\end{bmatrix*}. \label{eq123456}
\end{align}
The priority vector $q(\cdot)$  of the $K$ matrix in~\eqref{eq123456} is given as $q(\cdot)=1\mapsto \begin{bmatrix} \textcolor{cyan}{2} & \textcolor{cyan}{1} \end{bmatrix}$, $q(\cdot)=2\mapsto \begin{bmatrix} \textcolor{cyan}{1} & \textcolor{cyan}{5} \end{bmatrix}$, 
$q(\cdot)=3\mapsto \begin{bmatrix} \textcolor{cyan}{3} & \textcolor{cyan}{5} \end{bmatrix}$, 
$q(\cdot)=4\mapsto \begin{bmatrix} \textcolor{blue}{3} & \textcolor{blue}{7} & \textcolor{blue}{5} & \textcolor{blue}{8} \end{bmatrix}$, 
$q(\cdot)=5\mapsto \begin{bmatrix} \textcolor{blue}{3} & \textcolor{blue}{1} & \textcolor{blue}{3} & \textcolor{blue}{6} \end{bmatrix}$,
$q(\cdot)=6\mapsto \begin{bmatrix} \textcolor{blue}{7} & \textcolor{blue}{2} & \textcolor{blue}{6} & \textcolor{blue}{4} \end{bmatrix}$.  \\
We assume the attacker attacks the control blocks pertaining to the priority of $q(\cdot)\,=\,5$. Then, exploiting Algorithm~$4$, we obtain post-attack $N_{\text{final}}(K)$ as, $q(\cdot)=1\mapsto \begin{bmatrix} 0 & 0 \end{bmatrix}$, $q(\cdot)=2\mapsto \begin{bmatrix} \textcolor{black}{0} & \textcolor{black}{0} \end{bmatrix}$, 
$q(\cdot)=3\mapsto \begin{bmatrix} 3 & 5 \end{bmatrix}$, 
$q(\cdot)=4\mapsto \begin{bmatrix} 3 & 7 & 5 & 8 \end{bmatrix}$, 
$q(\cdot)=5\mapsto \begin{bmatrix} \textcolor{green}{3} & \textcolor{green}{1} & \textcolor{green}{3} & \textcolor{green}{6} \end{bmatrix}$,
$q(\cdot)=6\mapsto \begin{bmatrix} 7 & 2 & 6 & 4 \end{bmatrix}$.\footnote{The null matrix $\begin{bmatrix}0 & 0 \end{bmatrix}$ symbolizes the control blocks which are scarified to reroute the attacked information. It is illustrated in Fig.~\ref{121213} by dotted lines.}\\  
Notice that, as stated in Remark~\ref{rema123}, the attacked information with the priority level $q(\cdot)=5$ contains $4$ unit of information and the information is rerouted through the communication links with the priority level $q(\cdot)=1$ and $q(\cdot)=3$ respectively ($2$ unit each). An equivalent pictorial representation is presented in Fig.~\ref{121213}.

\begin{rem} 
{\rm The rerouting algorithms proposed in Algorithm~$2, 3$ and $4$, are general enough to tackle cyber-attacks for any class of structured gain $K\in\mathbb{R}^{m\times n}$ with large values of $m, n\in\mathbb{N}$. However, we adhere to the smaller dimensions of $K$ in example $1$ and $2$ in order not to blur the message of our work.}
\end{rem}
\vspace{0.2 cm}
\textbf{Step $3$: Structured $\mathcal{H}_2$ algorithm:} The structured $\mathcal{H}_2$ algorithm collects the post-attack $N_{\text{final}}(K)\in\mathbb{R}^{r_1\times r_2}$ from Step-$2$. $N_{\text{final}}(K)$ contains the information corresponding to the zero and non-zero control blocks, and allows us to construct the post-attack structural constraint $\Omega$ (say $\hat{\Omega}$). Then our objective boils down to solve the following optimization problem
\begin{equation}
\begin{aligned}
& \underset{K}{\text{minimize}}
& & J(K) \\
& \text{subject to}
& & K\,\circ\,I^c_{\hat{\Omega}}\,=\,0,
\end{aligned}      \label{eq3001}
\end{equation}
where $I^c_{\hat{\Omega}}\,=\,\mathbf{1}\,-\,I_{\hat{\Omega}}$, and $I_{\hat{\Omega}}$ is the structural identity of the subspace $\hat{\Omega}$ with its $ij^{th}$ entry as
\[[I_{\hat{\Omega}}]_{ij}\,:=\,\begin{cases}
      1 & \text{if $K_{ij}$ is a free variable} \\
       0 & \text{if $K_{ij}\,=\,0$ is required}.
      \end{cases}
\]
The augmented Lagrangian for~\eqref{eq3001} is evaluated as
\begin{align}
\mathcal{L}_\gamma (K,\Lambda)\,=\,J(K)\,+\,\text{trace}(\Lambda^{\top}(K\,\circ\,I^c_{\hat{\Omega}}))\,+\,\frac{\gamma}{2}\,||K\,\circ\,I^c_{\hat{\Omega}}||^2_F, \label{eq1001}
\end{align}
in which, $\frac{\gamma}{2}\,||K\,\circ\,I^c_{\hat{\Omega}}||^2_F$ is introduced to locally convexify the Lagrangian. (see~\cite{lin2011augmented},~\cite{bertsekas1999nonlinear} for the details). In~\eqref{eq1001} the penalty weight $\gamma$ is a positive scalar, while, $\Lambda\in\mathbb{R}^{m\times n}$ is the Lagrange multiplier. 

\begin{algorithm}[H]\label{a3}
\caption{Structured $\mathcal{H}_2$ algorithm}
\begin{algorithmic}[1]
\State \textit{Initialize}  $\Lambda^0\,=\,0$, and $\gamma^0\,>\,0$
\For {$(i\,=\,0,\,1,\,\cdots)$} 
\State For fixed $\Lambda^i$, minimize~\eqref{eq1001} with respect to unstructured $K$
\State Update $\Lambda^{i+1}\,=\,\Lambda^i\,+\,\gamma\,(K\,\circ\,I^c_{\hat{\Omega}})$ 
\State Update $\gamma^{i+1}\,=\,\alpha\,\gamma^i$ with $\alpha\,>\,1$
\State \textbf{until:} the stopping criteria $||K^i\,\circ\,I^c_{\hat{\Omega}}||\,<\,\epsilon$
\EndFor
\end{algorithmic}
\end{algorithm}

The optimization paradigm given in~\eqref{eq3001}, always yields a post-attack stabilizing control gain $K$; see~\cite[Remark 3]{lin2013design} for details.\\

\begin{rem}
{\rm It is noteworthy that Step~$1$ of our proposed algorithm is mainly carried out offline, while Step $2$ and $3$ are performed online after an attack occurs in the network. Step~$1$ can be executed at any time before the attack, and it doesn't depend on the characteristics of the attack model.}
\end{rem}
\vspace{0.2 cm}
\section{\large{Simulation Results}} \label{num}
\begin{figure}[!htb]
	\centering
		\includegraphics[width = 3.5 in, keepaspectratio]{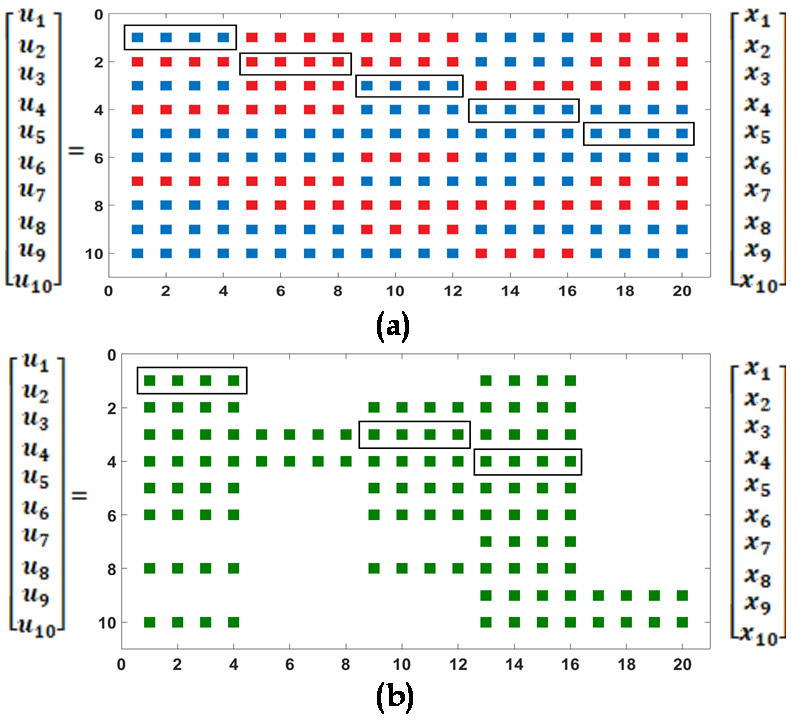}
	\caption{\small{(a) The sparsity pattern of the centralized LQR gain $K_c\in\mathbb{R}^{10\times 20}$. The ($\textcolor{red}{\bullet}$) denotes the attacked control blocks. The square box denotes the amount of information accumulated by each control block. (b) illustrates the post-attack sparsity pattern after applying the rerouting algorithm given in Section~\ref{mars}.}} \label{ce101}
\end{figure}

In this section, we present an academic example to illustrate the functionality of our proposed algorithm. We revisit a coupled dynamical system with $N\,=\,10$ sub-systems, as
\begin{align}
\dot{x}_i(t)\,=\,\sum_{j=1}^N A_{ij}\,x_i(t)\,+\,B_{ii}\,u_i(t)\,+\,W_{ii}\,d_i(t), \quad \forall i\,\in\,\left\lbrace 1,\cdots,N\right\rbrace, \label{eq100}
\end{align}
where, $x_i(t)\in\mathbb{R}^2$ is the state vector of the $i^{th}$ sub-system, while $d_i(t)\in\mathbb{R}^2$, and $u_i(t)\in\mathbb{R}$ are the corresponding disturbance and the control input. The overall system dynamics evolves as
\begin{align}
\dot{x}(t)\,=\,Ax(t)\,+\,Bu(t)\,+\,Wd(t), \label{eq1469}
\end{align} 
where the drift matrix $A\in\mathbb{R}^{20\times 20}$ is structured as
\begin{align*}
A\,=\,\begin{bmatrix}
A_{11} & A_{12} & \cdots & A_{1N}\\
A_{21} & A_{22} & \cdots & A_{2N}\\
\vdots & \vdots & \ddots & \vdots\\
A_{N1} & A_{N2} & \cdots & A_{NN}
\end{bmatrix}.
\end{align*} 
We consider $A$ to be a random matrix, in which, all the eigenvalues lie in the open left half of the complex plane, while the dominant eigenvalue is placed relatively close to the origin. In addition, we calculate the overall $B\in\mathbb{R}^{20\times 10}$ and $W\in\mathbb{R}^{20\times 20}$ given in~\eqref{eq1469}, as
\begin{align*}
B\,=\,B_{ii}\otimes I_{10}, \quad D\,=\,D_{ii}\otimes I_{10}, \quad \forall i\in \left\lbrace 1,2,\cdots,N\right\rbrace  
\end{align*}
where
\begin{align*}
B_{ii}\,=\,\begin{bmatrix}
10\\
0
\end{bmatrix}, \quad W_{ii}\,=\,\begin{bmatrix}
0.5 & 0\\
0 & 0.5
\end{bmatrix}.
\end{align*}
We select the design parameters as $Q\,=\,I_{20}$ and $R\,=\,10I_{10}$ respectively. Given the system dynamics~\eqref{eq1469}, we consider the control law $u(t)\,=\,K\,x(t)$, where $K\in\mathbb{R}^{10\times 20}$ is the feedback gain matrix. The gain matrix $K$ is designed based on the sparsity promoting algorithm~\eqref{eq5}.  First, we start minimizing~\eqref{eq5} for a small initial value of $\beta$. We consider the block sparsity structure with the weighted sum of the Frobenius norm given in~\eqref{eq6}, as the sparsity-promoting penalty function and set $G_{ij}\,=\,1/\left(||K_{ij}||_F+\varepsilon\right)$ with $\varepsilon\,=\,10^{-3}$. As stated earlier, the small (initial) value of $\beta$ yields a centralized LQR gain $K_c\in\mathbb{R}^{10\times 20}$, depicted in Fig.~\ref{ce101}~(a). Then, we perform an offline analysis based on Algorithm~$1$ to collect all the non-zero control blocks of $K\in\mathbb{R}^{10\times 20}$ and prioritize each of them. We do the prioritization in terms of the loss of closed-loop performance when the corresponding block of the feedback gain matrix is removed. We assume, an anonymous attacker has the knowledge of the priority of each control blocks of $K$ and attacks a set of communication links (denoted in \textcolor{red}{red} in Fig.~\ref{ce101}~(a)) to yield a poor closed loop response. Finally, an application of the re-routing algorithms evaluates the post-attack sparsity structure illustrated in Fig.~\ref{ce101}~(b).

\begin{table}[hb]
\begin{center}
\caption{Comparison between the pre and post-attack $J(K)$ values}\label{tabl}
\vspace{0.2 cm}
\begin{tabular}{cccc}
$J(K)$ & Before attack & Immediate after attack & After reroute \\\hline
Fig~\ref{ce101} & $0.47$ & $0.48$ & $0.475$ \\ \hline
Fig.~\ref{ce102} & $0.46$ & $0.47$ & $0.467$ \\
\hline
\end{tabular}
\end{center}
\end{table}
Notice that, initially the centralized LQR gain $K\,=\,K_c$ has $50$ control blocks, where each control block carries $4$ unit of information (placed inside the rectangle shown in Fig.~\ref{ce101}~(a),~(b)). The attacker attacks $22$ control blocks (denoted in \textcolor{red}{red} in Fig~\ref{ce101}~(a)) which are successfully rerouted via rest of the $28$ control blocks shown in~\ref{ce101}~(b), as expected.

\begin{figure}[!htb]
	\centering
		\includegraphics[width = 3.5 in, keepaspectratio]{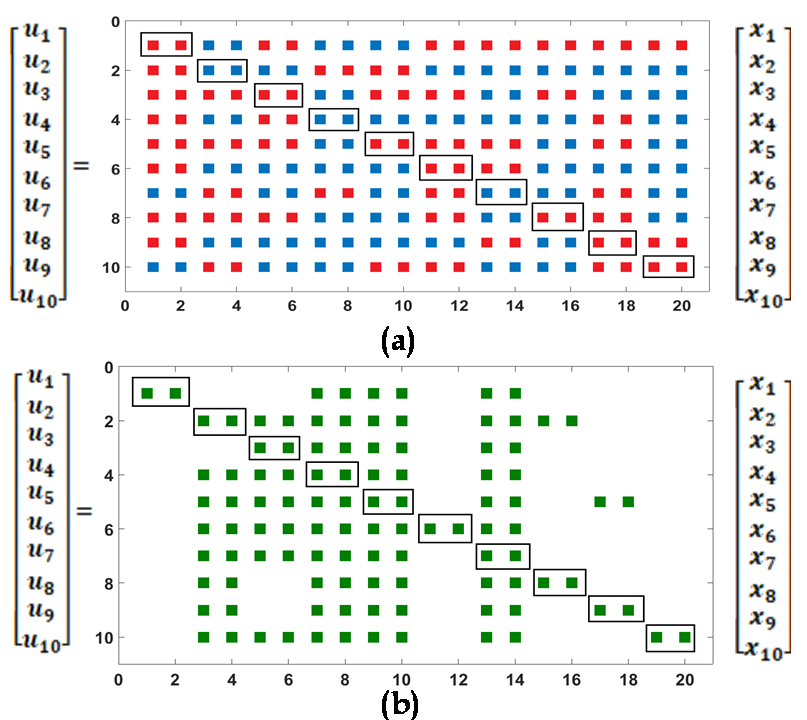}
	\caption{\small{(a) The sparsity pattern of the centralized LQR gains $K_c\in\mathbb{R}^{10\times 20}$. The ($\textcolor{red}{\bullet}$) denotes the control blocks which are attacked by the attacker. The square box signifies amount of information carried by each control block. In (b), we illustrate the post-attack sparsity pattern after applying rerouting algorithm given in Section~\ref{mars}.}} \label{ce102}
\end{figure}

Next, we simulate another numerical setup illustrated in Fig.~\ref{ce102}~(a),~(b). We assume the centralized LQR gain $K_c\in\mathbb{R}^{10\times 20}$ partitioned into $100$ control blocks, in which, each control block contains $2$ unit of information. Notice that, the attacker attacks $48$ control blocks (denoted in \textcolor{red}{red} in Fig.~\ref{ce102}~(a)). Then, by exploiting the rerouting algorithm the attacked information are safely rerouted via remaining unaffected (Lower priority) $52$ control blocks; see Fig.~\ref{ce102}~(b).   

In Table~\ref{tabl}, we present a comparative study of the $J(K)$ values corresponding to both the simulations stated above.\\

\begin{rem}
{\rm Notice that, the figures presented above in particular, Fig.~\ref{ce101} (b),~\ref{ce102} (b), illustrate the post-attack structural constraint, in which, we employ the structured $\mathcal{H}_2$ algorithm given in Algorithm~$5$, to evaluate the post-attack feedback gains and its corresponding $J(K)$ values. The post-attack (after rerouting) $J(K)$ values are documented in Table~\ref{tabl}, however, we exclude the post-attack feedback gain matrices due to the space limitation.}
\end{rem}
\vspace{0.2 cm}
\begin{rem}
{\rm In Table I, it seems that not a lot of control performance is lost when the DoS attack is happening, and also not a lot of performance is gained when the proposed countermeasure is applied to the attacked systems. This is occurred due to the choice of the system matrix $A$ given in Eq.~\eqref{eq1469}. Notice that, $A$ is a random matrix in which each element lies between $[0,1]$. A larger value of the system matrix also alter the control performances. However, as stated earlier, we adhere to simple examples to validate our contributions.}
\end{rem}


\section{\large{Concluding remarks}}
Given a continuous-time LTI system, this work highlights a new algorithmic strategy to alleviate cyber-attacks for LTI systems. We assume the anonymous attacker has the information corresponding to the structured feedback gain matrix and the priority of its control channels. Then based on the attack model, we develop a rerouting strategy, in which, after an attack, the higher priority communication data are rerouted through lower priority control channels. The priority of the control channels are assigned employing sparse optimization methods with sparsity promoting penalty functions. The rerouting functions allow us to determine the post-attack structural constraint, in which an application of the structured $\mathcal{H}_2$ algorithm evaluates the post-attack feedback gain. Future work involves analyzing the complexity of the proposed algorithms.

\ifCLASSOPTIONcaptionsoff
  \newpage
\fi

\small
\bibliographystyle{IEEEtranN}
\bibliography{diss}

\begin{thebibliography}{20}
\providecommand{\natexlab}[1]{#1}
\providecommand{\url}[1]{#1}
\csname url@samestyle\endcsname
\providecommand{\newblock}{\relax}
\providecommand{\bibinfo}[2]{#2}
\providecommand{\BIBentrySTDinterwordspacing}{\spaceskip=0pt\relax}
\providecommand{\BIBentryALTinterwordstretchfactor}{4}
\providecommand{\BIBentryALTinterwordspacing}{\spaceskip=\fontdimen2\font plus
\BIBentryALTinterwordstretchfactor\fontdimen3\font minus
  \fontdimen4\font\relax}
\providecommand{\BIBforeignlanguage}[2]{{%
\expandafter\ifx\csname l@#1\endcsname\relax
\typeout{** WARNING: IEEEtranN.bst: No hyphenation pattern has been}%
\typeout{** loaded for the language `#1'. Using the pattern for}%
\typeout{** the default language instead.}%
\else
\language=\csname l@#1\endcsname
\fi
#2}}
\providecommand{\BIBdecl}{\relax}
\BIBdecl

\bibitem[Ehrenfeld(2017)]{ehrenfeld2017wannacry}
J.~M. Ehrenfeld, ``Wanna{C}ry, cybersecurity and health information technology:
  A time to act,'' \emph{Journal of medical systems}, vol.~41, no.~7, p. 104,
  2017.

\bibitem[Fayi(2018)]{fayi2018petya}
S.~Y.~A. Fayi, ``What {P}etya/{N}ot{P}etya ransomware is and what its
  remidiations are,'' in \emph{Information Technology-New Generations}.\hskip
  1em plus 0.5em minus 0.4em\relax Springer, 2018, pp. 93--100.

\bibitem[Liang et~al.(2017)Liang, Weller, Zhao, Luo, and Dong]{liang20172015}
G.~Liang, S.~R. Weller, J.~Zhao, F.~Luo, and Z.~Y. Dong, ``The 2015 ukraine
  blackout: Implications for false data injection attacks,'' \emph{IEEE
  Transactions on Power Systems}, vol.~32, no.~4, pp. 3317--3318, 2017.

\bibitem[Shoukry and Tabuada(2016)]{shoukry2016event}
Y.~Shoukry and P.~Tabuada, ``Event-triggered state observers for sparse sensor
  noise/attacks,'' \emph{IEEE Transactions on Automatic Control}, vol.~61,
  no.~8, pp. 2079--2091, 2016.

\bibitem[Pasqualetti et~al.(2013)Pasqualetti, D{\"o}rfler, and
  Bullo]{pasqualetti2013attack}
F.~Pasqualetti, F.~D{\"o}rfler, and F.~Bullo, ``Attack detection and
  identification in cyber-physical systems,'' \emph{IEEE Transactions on
  Automatic Control}, vol.~58, no.~11, pp. 2715--2729, 2013.

\bibitem[Chen et~al.(2018)Chen, Kar, and Moura]{chen2018resilient}
Y.~Chen, S.~Kar, and J.~M. Moura, ``Resilient distributed estimation through
  adversary detection,'' \emph{IEEE Transactions on Signal Processing}, 2018.

\bibitem[Chen et~al.(2017)Chen, Kar, and Moura]{chen2017distributed}
------, ``Distributed estimation under sensor attacks,'' \emph{arXiv preprint
  arXiv:1709.06156}, 2017.

\bibitem[De~Persis and Tesi(2014{\natexlab{a}})]{de2014resilient}
C.~De~Persis and P.~Tesi, ``Resilient control under denial-of-service,''
  \emph{IFAC Proceedings Volumes}, vol.~47, no.~3, pp. 134--139, 2014.

\bibitem[De~Persis and Tesi(2014{\natexlab{b}})]{de2014non}
------, ``On resilient control of nonlinear systems under denial-of-service,''
  in \emph{53rd Annual Conference on Decision and Control (CDC), 2014
  IEEE}.\hskip 1em plus 0.5em minus 0.4em\relax IEEE, 2014, pp. 5254--5259.

\bibitem[Senejohnny et~al.(2015)Senejohnny, Tesi, and
  De~Persis]{senejohnny2015self}
D.~Senejohnny, P.~Tesi, and C.~De~Persis, ``Self-triggered coordination over a
  shared network under denial-of-service,'' in \emph{54th Annual Conference on
  Decision and Control (CDC), 2015 IEEE}.\hskip 1em plus 0.5em minus
  0.4em\relax IEEE, 2015, pp. 3469--3474.

\bibitem[Pasqualetti et~al.(2012)Pasqualetti, Bicchi, and
  Bullo]{pasqualetti2012consensus}
F.~Pasqualetti, A.~Bicchi, and F.~Bullo, ``Consensus computation in unreliable
  networks: A system theoretic approach,'' \emph{IEEE Transactions on Automatic
  Control}, vol.~57, no.~1, pp. 90--104, 2012.

\bibitem[Feng and Hu(2017{\natexlab{a}})]{feng2017}
Z.~Feng and G.~Hu, ``Distributed secure average consensus for linear
  multi-agent systems under dos attacks,'' in \emph{2017 American Control
  Conference (ACC)}.\hskip 1em plus 0.5em minus 0.4em\relax IEEE, 2017, pp.
  2261--2266.

\bibitem[Feng and Hu(2017{\natexlab{b}})]{feng2017distributed}
------, ``Distributed secure leader-following consensus of multi-agent systems
  under dos attacks and directed topology,'' in \emph{2017 IEEE International
  Conference on Information and Automation (ICIA)}.\hskip 1em plus 0.5em minus
  0.4em\relax IEEE, 2017, pp. 73--79.

\bibitem[Hota et~al.(2016)Hota, Clements, Sundaram, and
  Bagchi]{hota2016optimal}
A.~R. Hota, A.~A. Clements, S.~Sundaram, and S.~Bagchi, ``Optimal and
  game-theoretic deployment of security investments in interdependent assets,''
  in \emph{International Conference on Decision and Game Theory for
  Security}.\hskip 1em plus 0.5em minus 0.4em\relax Springer, 2016, pp.
  101--113.

\bibitem[Hota and Sundaram(2018)]{hota2018}
A.~R. Hota and S.~Sundaram, ``Interdependent security games on networks under
  behavioral probability weighting,'' \emph{IEEE Transactions on Control of
  Network Systems}, vol.~5, no.~1, pp. 262--273, 2018.

\bibitem[Shukla et~al.(2018)Shukla, Chakrabortty, and
  Duel-Hallen]{shukla2018cyber}
P.~Shukla, A.~Chakrabortty, and A.~Duel-Hallen, ``A cyber-security investment
  game for networked control systems,'' \emph{arXiv preprint arXiv:1810.00232},
  2018.

\bibitem[Lin et~al.(2013)Lin, Fardad, and Jovanovi{\'c}]{lin2013design}
F.~Lin, M.~Fardad, and M.~R. Jovanovi{\'c}, ``Design of optimal sparse feedback
  gains via the alternating direction method of multipliers,'' \emph{IEEE
  Transactions on Automatic Control}, vol.~58, no.~9, pp. 2426--2431, 2013.

\bibitem[Lin et~al.(2011)Lin, Fardad, and Jovanovic]{lin2011augmented}
F.~Lin, M.~Fardad, and M.~R. Jovanovic, ``Augmented lagrangian approach to
  design of structured optimal state feedback gains,'' \emph{IEEE Transactions
  on Automatic Control}, vol.~56, no.~12, pp. 2923--2929, 2011.

\bibitem[Zhou et~al.(1996)Zhou, Doyle, Glover, et~al.]{zhou1996robust}
K.~Zhou, J.~C. Doyle, K.~Glover \emph{et~al.}, \emph{Robust and optimal
  control}.\hskip 1em plus 0.5em minus 0.4em\relax Prentice hall New Jersey,
  1996, vol.~40.

\bibitem[Bertsekas(1999)]{bertsekas1999nonlinear}
D.~P. Bertsekas, \emph{Nonlinear programming}.\hskip 1em plus 0.5em minus
  0.4em\relax Athena scientific Belmont, 1999.

\end{thebibliography}
\end{document}